\catcode`\@=11

\newif\if@fewtab\@fewtabtrue
{\count255=\time\divide\count255 by 60
\xdef\hourmin{\number\count255}
\multiply\count255 by-60\advance\count255 by\time
\xdef\hourmin{\hourmin:\ifnum\count255<10 0\fi\the\count255}}
\def\ps@draft{\let\@mkboth\@gobbletwo
    \def\@oddhead{}
    \def\@oddfoot
     {\hbox to 7 cm{\tiny \versionno \hfil}\hskip -7cm\hfil\rm\thepage \hfil}
    \def\@evenhead{}\let\@evenfoot\@oddfoot}

\def\draftcite#1{\ifnum\draftcontrol=1#1\else{}\fi}

\def\@bibitem#1{\item\hskip -3cm \hbox to 2cm
{\hfil {\footnotesize\bf\draftcite{#1}}}\hskip 1cm
\if@filesw \immediate\write\@auxout
       {\string\bibcite{#1}{\the\value{\@listctr}}}\fi\ignorespaces}

\def\nsection#1{\section{#1}\setcounter{equation}{0}}

\global\def\draftcontrol{0}
\catcode`\@=12

\documentstyle[12pt]{article}

 \def\yes{yes }
\ifx\yesno\yes 
\else  \fi
 \read-1 to\yesno
\ifx\yesno\yes 
\font\tendl=msym10  scaled \magstep1
\font\sevendl=msym7 scaled \magstep1
\font\fivedl=msym5 scaled \magstep1
\newfam\dlfam \def\dl{\fam\dlfam\tendl}
\textfont\dlfam=\tendl \scriptfont\dlfam=\sevendl
\scriptscriptfont\dlfam=\fivedl
\else \let\dl=\bf
\fi

\def\ifundefined#1{\expandafter\ifx\csname#1\endcsname\relax}
\makeatletter \ifundefined{new@mathgroup} {} \else
    \new@mathgroup\sffam
    \new@internalmathalphabet\mathsf\sffam{cmss}{m}{n}
    \def\psf{\fontfamily\sfdefault \fontseries\default@series
        \fontshape\default@shape\selectfont\mathsf}
\fi

\def\citen#1{\if@filesw \immediate\write \@auxout {\string\citation{#1}}\fi%
\@tempcntb\m@ne \let\@h@ld\relax \def\@citea{}%
\@for \@citeb:=#1\do {\@ifundefined {b@\@citeb}%
    {\@h@ld\@citea\@tempcntb\m@ne{\bf ?}%
    \@warning {Citation `\@citeb ' on page \thepage \space undefined}}%
    {\@tempcnta\@tempcntb \advance\@tempcnta\@ne
    \setbox\z@\hbox\bgroup\ifcat0\csname b@\@citeb \endcsname \relax
    \egroup \@tempcntb\number\csname b@\@citeb \endcsname \relax
    \else \egroup \@tempcntb\m@ne \fi \ifnum\@tempcnta=\@tempcntb
    \ifx\@h@ld\relax \edef \@h@ld{\@citea\csname b@\@citeb\endcsname}%
    \else \edef\@h@ld{\hbox{--}\penalty\@highpenalty
    \csname b@\@citeb\endcsname}\fi
    \else \@h@ld\@citea\csname b@\@citeb \endcsname \let\@h@ld\relax \fi}%
\def\@citea{,\penalty\@highpenalty\hskip.13em plus.13em minus.13em}}\@h@ld}
\def\@citex[#1]#2{\@cite{\citen{#2}}{#1}}%
\def\@cite#1#2{\leavevmode\unskip\ifnum\lastpenalty=\z@\penalty\@highpenalty\fi%
  \ [{\multiply\@highpenalty 3 #1%
  \if@tempswa,\penalty\@highpenalty\ #2\fi}]}   %
\makeatother 

\def\A             {Algebra}
\def\abar          {\bar A}
\def\ag            {A\gt g}
\def\amu           {A_\mu}
\def\amutau        {A_\mu^{(\tau)}}
\def\anull         {\mbox{$A$}\hsp{-.4}\raisebox{.71em}{$\scriptstyle\circ$}}
\newcommand{\andauthoretc}[5]{\vskip1.2mm \centerline{and}{}\vskip2mm
                   \centerline{\sc #1}\vskip2mm
                   \centerline{#2}\vskip.2mm \centerline{#3}\vskip.2mm
                   \centerline{#4}\vskip.2mm \centerline{#5}}
\newcommand{\authoretc}[5]{\centerline{\sc #1}\vskip2mm
                   \centerline{#2}\vskip.2mm \centerline{#3}\vskip.2mm
                   \centerline{#4}\vskip.2mm \centerline{#5}}

\def\atau          {A_{}^{(\tau)}}

\def\bara          {\mbox{$\bar A$}}
\def\be            {\begin{equation}}
\def\bfe           {{\bf1}}
\def\cala          {\mbox{${\cal A}\,$}}

\def\calb          {\mbox{${\cal B}$}}
\def\calg          {\mbox{${\cal G}$}}
\def\calgo         {\mbox{${\cal G}_0$}}
\def\calgog        {\mbox{${\cal G}_0^{(\bar g)}$}}

\def\calm          {\mbox{${\cal M}$}}

\def\calo          {\mbox{$\cal O$}}
\def\cals          {\mbox{$\cal S$}}
\def\calsa         {\mbox{${\cal S}_{\!A}$}}
\def\calsab        {\mbox{${\cal S}_{\!\bar A}$}}
\def\calsnull      {\mbox{${\cal S}^{}_{\!A}\hsp{-.41}\raisebox{.19em}
                   {$\scriptscriptstyle\circ$}$}}

\def\cmu           {c_\mu}
\def\complex       {{\dl C}}
\def\confi         {configuration}

\long\def\del#1    \enddel{}

\def\delo          {\mbox{$\nabla_{\!\!\bar A}^{}$}}
\def\delom         {\mbox{$\nabla_{\!\!A}^{}$}}
\def\deloms        {\mbox{$\nabla_{\!\!A}^*$}}
\def\delos         {\mbox{$\nabla_{\!\!\bar A}^*$}}
\def\difeq         {differential equation}
\def\dmu           {\partial_\mu}
\def\dnu           {\partial_\nu}
\def\dmU           {\partial^\mu}
\def\dmugg         {\partial_\mu g\,g^{-1}}
\def\dmUgg         {\partial^\mu g\,g^{-1}}

\def\ee            {\end{equation}}
\def\eE            {{\rm e}}

\newcommand{\erf}[1]{(\ref{#1})}
\def\F             {\Phi_A^{}}
\def\FA            {\Phi_{\bar A;A}^{}}
\def\FAA           {\Phi_{\bar A;\bar A}^{}}
\def\FAC           {\Phi_{\bar A;C}^{}}
\def\FAnull        {\Phi^{}_{\!A;A}\hsp{-1.15}\raisebox{.19em}
                   {$\scriptscriptstyle\circ$}\hsp{.79}}
\def\findim        {finite-di\-men\-si\-o\-nal}

\def\fmd           {fundamental modular domain}
\def\fmu           {a_\mu}
\def\fmU           {a^\mu}
\def\fnu           {a_\nu}
\def\FP            {Faddeev\hy Popov}
\def\fp            {\mbox{$\Delta_{\rm FP} $}}
\def\fpbara        {\mbox{$\Delta_{\rm FP}^{(\bar A,A)} $}}
\def\futnote#1     {\footnote{~#1}\ }
\def\g             {{\sf g}}
\def\gammA         {\sigma}
\def\gbar          {{\bar g}}
\def\gc            {g_{\rm c}}
\def\gg            {g^{-1}}

\def\go            {g_0^{}}
\newcommand{\goorb}[1]{{\cal Q}_{\!\displaystyle #1}}
\def\gome          {g_0^{-1}}
\newcommand{\gt}[1] {^{#1}}
\def\gtrafo        {gauge transformation}

\def\hg            {\gt g}

\newcommand{\hsp}[1] {\mbox{\hspace{#1 em}}}
\def\hy            {$\mbox{-\hspace{-.66 mm}-}$}

\def\ii            {{\rm i}}
\def\infdim        {infi\-nite-di\-men\-si\-o\-nal}
\def\Infdim        {Infi\-nite-dimen\-sio\-nal}
\def\ixxx          {\int_{\cal C}\!{\rm d}^3x\,}
\def\ixxxx         {\int\!{\rm d}^4x\,}
\def\ixxxxm        {\int_M \!{\rm d}^4x\,}
\def\ixxxxM        {\mbox{\Large$\int$}_{M}^{}{\rm d}^4x\,}

\def\kon           {\mbox{$\Omega$}}
\def\konA          {\mbox{$\Omega_{\bar A}$}}
\def\L             {\mbox{$\Lambda$}}
\def\LA            {\mbox{$\Lambda_{\bar A}$}}
\def\La            {\mbox{$\breve\Lambda$}}

\long\def\labl#1   {\label{#1}\ee}

\def\Lanul         {\mbox{$\hat\Lambda^{}_{\!A}\hsp{-.41}\raisebox{.19em}
                   {$\scriptscriptstyle\circ$}$}}
\def\Lanull        {\mbox{$\breve\Lambda^{}_{\!A}\hsp{-.41}\raisebox{.19em}
                   {$\scriptscriptstyle\circ$}$}}
\def\LB            {\mbox{$\Lambda_{B}$}}
\def\LG            {\mbox{$\Lambda_{\bar A}/{\cal S}_{\!\bar A}$}}
\def\LGb           {\mbox{$\breve\Lambda_{\bar A}/{\cal S}_{\bar A}$}}

\def\Lnull         {\mbox{$\Lambda^{}_{\!A}\hsp{-.41}\raisebox{.19em}
                   {$\scriptscriptstyle\circ$}$}}

\def\LO            {\mbox{$\tilde\Lambda$}}

\def\one           {\mbox{\small $1\!\!$}1}
\def\onehalf       {\mbox{$\frac12$}}

\def\pb            {{\cal P}_B}
\def\pc            {{\cal P}_C}
\def\PSI           {\Gamma}
\def\Q             {Quantum }

\long\def\query#1{\hskip 0pt{\vadjust{\everypar={}\small\vtop to 0pt{\hbox{}%
     \vskip -13pt\rlap{\hbox to 48.9pc{\hfil{\vtop{\hsize=8pc\tolerance=6000%
     \hfuzz=.5pc\rightskip=0pt plus 3em\noindent#1}}}}\vss}}}}%
\long\def\rank#1   {\mbox{rank}\,#1}
\def\reals         {{\dl R}}
\def\realspluso    {{\dl R}_{\geq0}}
\def\rep           {representation}

\newcommand{\restr}[1] {\raisebox{-.25em}{$|$}\raisebox{-.45em}{$\scriptstyle
#1$}}
\def\s             {\mbox{$\sigma$}}

\def\semitimes     {\begin{picture}(7.2,8)\put(1.55,0.32){\line(0,1)
                   {4.51}}\put(0,0){$\times$} \end{picture} }

\def\smi           {smooth inner }
\def\st            {space-time}
\def\taudr         {\mbox{$\tau^3$}}

\def\tauo          {\tau}
\def\tr            {{\rm tr}\,}

\def\twodim        {two-di\-men\-si\-o\-nal}

\def\wrt           {with respect to }
\def\xo            {\mbox{$x_0$}}
\def\xO            {{x_0}}
\def\YM            {Yang\hy Mills}
\def\zet           {{\dl Z}}

\begin{document} 

\begin{flushright}  {~} \\[-23 mm] {\sf hep-th/9404059} \\
{\sf HD-THEP-94-11}  \\ {\sf NIKHEF-H/94-11} \\[1 mm]{\sf April 1994}
\end{flushright} \vskip 6mm
\begin{center}

{\Large{\bf{ON THE CONFIGURATION SPACE}}} \vskip 0.3cm

{\Large{\bf OF GAUGE THEORIES}} \end{center}
\vskip 9mm{} {\authoretc{\ J\"urgen Fuchs\ $^\pounds$}
    {NIKHEF-H}{Kruislaan 409} {NL -- 1098 SJ~~Amsterdam}{}
    \andauthoretc{\ Michael G.\ Schmidt}{Institut f\"ur
       Theoretische Physik} {Philosophenweg 16} {D -- 69120~~Heidelberg} {}}
    \andauthoretc{\ Christoph Schweigert} {NIKHEF-H\,/\,FOM}{Kruislaan 409}
    {NL -- 1098 SJ~~Amsterdam}{}


\vskip 1cm \begin{quote}{\bf Abstract}.
We investigate the structure of the configuration space of gauge
theories and its description in terms of the set of absolute minima of
certain Morse functions on the gauge orbits. The set of absolute minima
that is obtained when the background connection is a pure gauge
is shown to be isomorphic to the orbit space of the pointed gauge
group. We also show that the stratum of irreducible orbits is
geodesically convex, i.e.\ there are no geometrical obstructions to
the classical motion within the main stratum.
An explicit description of the singularities of the configuration space
of SU(2) theories on a topologically simple space-time and on the lattice
is obtained; in the continuum case we find that the singularities are
conical and that the singular stratum is isomorphic to a $\zet_2$-orbifold
of the configuration space of electrodynamics. \end{quote} 
---------------------\\[1 mm]
{\small $^\pounds$ \ Heisenberg fellow} \vskip 1.3cm

\nsection{The space of gauge orbits}

For both phenomenological and conceptual reasons,
gauge theories play a prominent r\^ole in physics. A basic object of
interest is the full configuration space of a gauge theory; in the context
of quantization by path integrals this space is relevant simply because
it is the path integration domain, while in
the Hamiltonian formalism one has to deal with wave functionals that
are defined as functionals over the whole configuration space.
The most impressing successes of gauge theories emerged in the context
of perturbation theory. In order to analyze with similar success
various nonperturbative
aspects of gauge theories, such as the confinement problem,
a detailed understanding of the geometry of the configuration space,
including its global features, will presumably be necessary.

In this note we comment on the configuration space of gauge theories and on
its description in terms the absolute minima of certain Morse functions.
More specifically, we consider pure \YM\ theories. Such theories are
already interesting in themselves; moreover,
for more complicated theories, such as \YM\ coupled to matter without
or even with spontaneous symmetry breaking, the configuration space
contains the configuration space of pure \YM\ theory as a subspace,
so that a detailed knowledge of the latter is again compulsory.
The action functional of pure \YM\ theory is given by
  \be 4{\rm g}^2\,S_{\rm YM}[A] = |F|^2 \equiv
  \int_M{\rm d}^4x\, \tr (F_{\mu \nu}F^{\mu\nu}) \,.  \labl{lagr}
Here g is the coupling constant, $A=A_\mu{\rm d}x^\mu$ is a
connection 1-form which takes values in the
Lie algebra $\g={\sf Lie}(G)$ of a finite-dimensional compact nonabelian Lie
group $G$ (called the {\em structure group\/} of the theory), and
$F_{\mu \nu}= \dmu A_\nu - \dnu A_\mu + [A_\mu,A_\nu] $ (the \YM\ field
strengths) are the components of the curvature 2-form $F={\rm d}A+A\wedge A$.
In \erf{lagr} we also introduced a norm $|\cdot|$ on the space of
equivariant $p$-forms over \st\ $M$ with values in \g; this norm is
defined as $|B|^2=(B,B)$, where $(\cdot\,,\cdot)$ is a scalar product on that
space which is defined by contracting group indices with the Killing form of
\g\ and integrating over \st,
  \be  (B,C) := \int_M \tr (B \wedge * C)  \, . \labl{skp}

To be able to make the path integral quantization well-defined, we take
the \st\ manifold to have euclidean signature. In the present article
we take $M$ to be the four-sphere $M = S^4$ which is the conformal
one point compactification $S^4=\reals^4\cup\{\infty\}$ of $\reals^4$.
This compactification corresponds to imposing suitable asymptotic
conditions at infinity; it
has the additional benefit of giving \st\ a finite volume.
Also note that the choice $M=S^4$ is familiar from the discussion
of instanton effects \cite{atjo}. In particular, the choice of
asymptotic conditions for the connections $A$ corresponds to the
choice of a specific (isomorphism class of) principal bundle $P$ with fiber
$G$ over $M$. For $M=S^4$ and $G$ a simple group, $P$ is labelled by
the instanton number $k\in\zet$; in the sequel we consider a fixed
principal bundle $P$, and hence the instanton number has a fixed value.

In a Hamiltonian formulation, the action \erf{lagr} leads to first
class constraints. These imply that in order to obtain a consistent
system with unique time evolution, we have to describe the system not
on the space \cala\ of all connections, but rather to factor out the
{\em gauge group\/} \calg. (By definition, \calg\ is the group of automorphisms
of the principal bundle $P$ which get projected to the identity on $M$.
Equivalently, \calg\ can be described by sections of the `gauge bundle'
$P \times_G G$ over $M$, where the fiber $G$ is endowed with the adjoint action
of $G$ on itself. In the latter description, locally any $g\in\calg$ is just
a map from \st\ to the finite-dimensional structure group $G$.)
Thus the true configuration space is the space
  \be  \calm := \cala / \calg = \{\calo_A\mid A\in\cala\} \labl m
of gauge orbits
  \be  \calo_A:=\{B\in\cala\mid B=\ag\,\ {\rm for\ some}\,\ g\in\calg\} \,.
  \labl o
Here for any $g\in\calg$
  \be  \ag=g^{-1}A\,g +g^{-1}\,{\rm d}g  \labl g
is the gauge transform of the connection $A$.
The norm introduced in \erf{lagr} is invariant under the action \erf g of
the gauge group, and hence in particular $S_{\rm YM}[A]= S_{\rm YM}[\ag]$
so that $S_{\rm YM}$ is well-defined on \calm.

While \cala\ is an affine space, the space \calm\ has a rather
complicated structure; in particular for generic choices of the
underlying \st\ manifold the topology of \calm\ is non-trivial \cite{sinG}.
For various purposes (such as the construction of a well-defined Feynman\hy Kac
path integral \cite{asmi}) one needs on \calg\ not only the group structure,
but also a topological structure, so that it becomes an \infdim\ Lie group;
technically, this can be accomplished by completion
\wrt a suitable Sobolev norm on \calg\ \cite{naRa,mivi,sefr,koro}.
It follows that when considering a non-compact \st,
in particular all constant gauge transformations
except for the unit element \bfe\ (defined by $\bfe(x)=\one \in G$ for all
$x\in M$) have to be excluded.
In our approach this is avoided by compactifying \st\ to $S^4$.

When coupling the \YM\ theory to chiral fermions, the non-trivial topology
of the orbit space \calm\ is the source of the chiral anomaly. Namely,
it leads to a gauge dependence of the phase of the fermion determinant so
that the determinant is well-defined only over \cala\ and not over \calm.
In other words, in the case of anomaly free theories the relevant bundle,
the determinant bundle over \calm, is trivial, while for a theory with a
chiral anomaly it is non-trivial and does not
admit global sections like the fermion determinant \cite{algi,hkprs}.

The rest of this paper is organized as follows. The next three sections
deal with various aspects of gauge fixing.
We work with background gauges in which these aspects are particularly
transparent and which in the geometric setting serve as a
natural starting point. In section 2 we introduce the
Gribov region and recall its description in terms of Morse functionals.
If the background connection is reducible, then some aspects of the
gauge fixing have to be treated with special care; this is described
in section 3. In particular, it is explained how \calm\ acquires the
structure of a stratified variety. In section 4 we analyze the structure
of the fundamental modular domains, which have the property that they
are isomorphic to the orbit space up to boundary identifications.
When the background connection is a pure gauge, then the set of absolute
minima of the Morse functionals is isomorphic to the orbit space of
the pointed gauge group; this is proven in section 5, while in
section 6 we show that the main stratum
is geodesically convex. The final section 7 contains a detailed
description of the configuration space of SU(2) \YM\ theory on $S^4$; we
show that in this case \calm\ decomposes in three strata, and that the strata
corresponding to reducible connections form conical singularities and
can be described as an
orbifold of the configuration space of electrodynamics; we also point out
the existence of reducible connections in the lattice version of the theory.

Among the central objects of our interest are the reducible connections.
The presence of reducible connections implies that the configuration space
\calm\ of \YM\ theories is {\em not} a manifold, but contains singularities;
the singular strata of \calm\ are formed by the orbits of reducible
connections.
While the physical implications of these singularities are still unclear,
it is worth emphasizing that they provide an intriguingly rich
structure that certainly deserves further investigation. In this
paper we obtain several new results which allow for a
more detailed description of the singular strata. We are confident that
these will ultimately be helpful in relating the singularities of the
configuration space to physical effects.

\nsection{Gauge fixing}

The non-trivial topological properties of the orbit space \calm\ make
it difficult to describe \calm\ directly. Accordingly for various purposes
it is necessary to resort to the covering space \cala\ of \calm\ and
identify in \cala\ an appropriate region that upon dividing out \calg\
projects bijectively on some open subset of \calm. As we will see below,
one can find subsets of \cala\ which are isomorphic to \calm\
modulo certain  boundary identifications; any such subset of \cala\ will
be called a {\em \fmd\/} for \calm.

A necessary requirement for identifying a fundamental
region is to `fix a gauge' in \cala, i.e.\ to choose, in a continuous
manner, representatives out of the gauge orbits \calo. The gauge fixing
we are going to use is implemented by means of a background gauge, which
is the most natural gauge condition from a geometric point of view.
That is, we choose
some arbitrary, but fixed, connection  $\abar\in\cala$ as the background,
and keep only those connections that belong to the affine subspace
  \be \PSI\equiv\PSI_{\!\abar}^{} := \{A \in \cala \mid \delos(A -\abar) =
  0 \} . \labl{psi}
Here \delom\ denotes the covariant derivative \wrt $A$, which acts
on one-forms $B$ as \,$\delom(B)={\rm d}B+[A,B]$.

The restriction to $\PSI$ certainly reduces the `number' of degrees of freedom
of the system (which remains infinite, of course). However, as it turns
out, the gauge fixing is not complete, i.e.\ generically more than one
element of a gauge orbit \calo\ satisfies the gauge condition (this is
not an artefact specific to background gauges, but in fact happens
for any continuous gauge fixing procedure \cite{sinG,bavi}).
Thus the subspace $\PSI$ of \cala\ together with the natural projection
to orbits fails to provide a coordinate system of \calm, i.e.\ $\PSI$
cannot yet serve as a \fmd. This can be made more precise; namely,
gauge copies appear at least outside a certain convex subset
$\kon\equiv\kon_{\abar}$ of $\PSI$ that can be characterized \cite{zwan1}
as the set of all minima of the functionals
  \be  \F\equiv\FA:\ \calg\rightarrow\realspluso\,, \qquad
  \F[g] := |\ag - \abar|^2   \labl{mors}
for $A\in\cala$.
 (The minima of $\F$ carry information about the topology of the
gauge orbit $\calo_A$ through $A$. Accordingly, the functional $\F$ is
called a Morse function \cite{vanb3}.)
\kon\ is known as the {\em Gribov\/} \cite{grib} {\em region}, and
its boundary $\partial\kon$ as the {\em Gribov horizon}. One can show that
any gauge orbit \calo\ intersects \kon\ at least once \cite{zwan1,dezw3},
and that \kon\ is is convex and bounded \cite{zwan1}.

To establish some equations that characterize the set \kon\ more
concretely, it is convenient to look \cite{sefr} at a local
one parameter family $g_t =\eE^{t w}$ of elements
of the gauge group \calg\ (thus $w=w(x)$ takes values in the Lie algebra
\g\ of $G$). For the first variation one finds
  \be \frac{{\rm d}}{{\rm d}t}\, \F[g_t] \restr{t=0} = 2\,(A-\bar A,\delom w)
  =-2\,(\deloms(A-\bar A) , w) \,, \ee
(the last expression is, in a first step, to be understood in the sense
of weak derivatives). Thus the first variation of $\F$ is
  \be \frac{\delta \F}{\delta g} = -2\,\deloms(A-\bar A)
  = -2\,\delos(A-\bar A) \,.  \ee
Thus in particular the definition \erf{psi} of $\PSI$ guarantees that
the points $A\in\PSI$ are characterized by the property that
$g_t\restr{t=0}\equiv \bfe\in\calg$ is a stationary point of $\F$.

Similarly we obtain for the second variation
  \be \frac{{\rm d^2}}{{\rm d} t^2}\, \Phi[g_t] \restr{t=0} =
  2\,(\delom w, \delom w)+ 2\,(A - \bar A, \delom w)= 2\,(\delo w,\delom w) =
   - 2\,(w, \delos\delom w) \,, \labl+
where again the last equality is to be read in the weak sense.
The Hessian of the variation is thus given by the \FP\ operator
  \be  \fp\equiv\fpbara:= - \delos \delom \,.  \labl{fp}
Note that
  \be  \deloms \delo = \delos \delom  \labl{gg}
for all $A \in \PSI$, and hence $\fp$ is symmetric.
The determinant of \fp\ is the usual \FP\ gauge fixing determinant, which
is closely related \cite{bavi0} to a natural Riemannian metric on \calm.
According to \erf+ the positivity of the \FP\ operator \fp\ ensures that
the connection $A$ possesses the property that $\bfe\in\calg$ is a
minimum of $\F$.

\nsection{Reducible connections}

While the previous statements are valid for any choice of the
background \bara, some of the properties of $\PSI$ and \kon\ do
depend on this choice. Namely, we have to distinguish between
irreducible and reducible backgrounds.
Here by an {\em irreducible\/} connection $A$ we mean
 \futnote{In the literature the term `irreducible' is sometimes (see e.g.\
\cite{bavi}) only used for connections with maximal holonomy;
such connections have in particular a trivial stabilizer.}
a connection for which the stabilizer (or isotropy subgroup)
  \be  \calsa=\{g\in\calg \mid \ag=A\}  \ee
of the action of the gauge group is trivial, i.e.\ equal to the
center $Z(G)$ of the structure group $G$ (clearly,
the constant gauge transformations corresponding
to the elements of $Z$ are in the stabilizer of any connection).
It is a standard result in the theory of principal bundles that the stabilizer
group $\calsa$ of any connection $A$ is isomorphic to the centralizer of
the holonomy group $H\equiv H^{}_A$ of $A$ relative to the finite-dimensional
structure group $G$ \cite[Lemma 4.2.8]{DOkr}.
 (The centralizer of a subgroup $\tilde G$ of $G$ consists of all
elements of $G$ that commute with all elements in $\tilde G$. It is
again a subgroup of $G$ and contains $Z(G)$ as a subgroup.)
For any $A\in\cala$, the holonomy group $H$ is a Lie subgroup of the \findim\
structure group $G$ \cite[p.\ 132]{DOkr}, and hence any stabilizer $\calsa$
is isomorphic to a closed subgroup of $G$ \cite{koro}.

It is known that the set \calm\ of orbits of all connections forms a
connected, separable and metrizable (and hence in particular Hausdorff)
topological space, and that this space has the structure of a stratified
variety, i.e.\ as a set is the disjoint sum of certain strata which are
smooth manifolds; the set of orbits of reducible connections is a closed subset
of this variety which is nowhere dense \cite{koro,sinG2,koSa,hkprs2}.
It has also been shown that
each stratum carries the structure of a Hilbert manifold, i.e.\ an
\infdim\ $C^\infty$ manifold modelled on a Hilbert space,
and its pre-image in \cala\ is a smooth $G$-invariant submanifold of
\cala \cite{naRa,mivi,koro,koSa}.

Each stratum of the variety \calm\ consists of all orbits of connections
that are of the same `symmetry type' in the sense \cite{koSa} that their
stabilizers are conjugate subgroups of \calg. (In particular, the stabilizers
of
all connections belonging to a fixed stratum are all isomorphic when
considered as abstract Lie groups. For the so-called main
stratum which consists of the orbits of irreducible connections, the
stabilizer is just the set of constant gauge transformations with values in
$Z(G)$.) Thus the strata of \calm\ can be labelled by conjugacy classes of
closed subgroups of the gauge group \calg\ that are isomorphic to
Lie subgroups of the \findim\ structure group $G$.
The set of such classes, and hence the set of strata of \calm, is
countable \cite{koro,koSa}. (Depending on the
topological properties of the \st\ $M$ and of $G$, the number of strata
may be finite; this will be the situation in the case of $M=S^4$ and
$G=\,$SU(2) that will be treated in section \ref{secsu2}.) The inclusion
relation among conjugacy classes of stabilizers in \calg\ induces
a partial ordering of the stabilizers. As the latter are used to label the
strata, this carries over to a partial ordering of strata.
One can show \cite{koro} that the stratum with stabilizers conjugated to
some given stabilizer \cals\ is dense in the union of all strata that
have stabilizers containing \cals. In particular the main stratum, which
has trivial stabilizer, is dense in the configuration space \calm, so that
the singular strata can be approximated arbitrarily well by irreducible
connections.

For practical purposes one is often interested in {\em reducible\/}
(i.e., non-irreducible) background connections.
(In section \ref{secsu2} below we will describe in some detail the
reducible connections in the case where the structure group is $G=\,$SU(2).)
In particular, the background $\abar=\anull$, where \anull\ stands for
the `vacuum', i.e.\ the configuration
  \be  \anull_{\mu}(x)\equiv0 \,, \ee
is reducible, and of course many calculations are simplified by this choice
of background. Now when fixing the gauge around a reducible background,
various subtleties have to be taken into account.

First of all, only if the background connection \bara\ is irreducible,
then \cite{daVi} any two connections $A,\,B$ ($A\neq B$) in $\PSI$ that are
sufficiently close to \bara\ in the topology induced by the scalar product
\erf{skp} belong to distinct gauge orbits $\calo_A\neq\calo_B$. In contrast,
for reducible background, $\PSI$ does not have this property. Further, for
irreducible background \bara, the Gribov region \kon\ can be defined as the
subspace of $\PSI$ on which the \FP\ operator $\fp$ is positive,
  \be  \kon\equiv\kon_{\abar}= \{A\in\PSI_{\!\abar}\mid (A, \fp A) \geq0\}
  \,; \labl{pos}
in particular, the Gribov horizon $\partial\kon$ can be characterized
by the vanishing of the \FP\ determinant $\,\det(\fp)$.  On the contrary,
for reducible connections there always exists \cite[p.\ 132]{DOkr} a
covariantly constant global section $\s$ of the vector bundle $P\times_G^{}\g$
(the `adjoint bundle') over $M$, where $G$ acts on $\g={\sf Lie}(G)$ in the
adjoint representation, i.e.\ a global section satisfying $\delo \s =0$.
To derive the existence of such a section, we have to keep in mind that
$g$ is a section of the bundle $P \times_G G$, and to note that
the gauge transformation \erf g can also be written in the form
  \be  A^g = A + g^{-1} \nabla_{\!\!A}\, g  \,. \labl{alt}
When considered for the background \bara, this equation
shows that we can describe the elements of the stabilizer
$\cals_{\!\abar}$ by covariantly constant sections in $P\times_GG$.
Now any stabilizer is a finite-dimensional Lie subgroup of \calg;
in particular, if the stabilizer is non-trivial (so that the dimension
of this Lie subgroup is not zero), we can consider a smooth
family $\{g_t(x)\mid t\in (-1,1) \}$ of such sections that are connected
to the unit element \bfe\ of \calg, i.e.\ $g_{t=0}^{}(x) =\one$ for all
$x\in M$. By differentiating this with respect to $t$, we see that
$\s:= \frac{{\rm d}}{{\rm d}t} g_t(x)\restr{t=0}$ is a
covariantly constant section of $P \times_G \g$.

Because of the symmetry \erf{gg} of the \FP\ operator, such a section \s\ obeys
  \be \delos \delom \s = \deloms \delo \s = 0 , \ee
i.e.\ $\s$ is in the kernel of \fp, from which it
follows that $\,\det(\fp)= 0$. Therefore in the case of reducible
backgrounds the Gribov horizon $\partial\kon$ cannot be characterized as the
set of all points of $\PSI$ where the \FP\ determinant vanishes. However,
it is still possible to describe $\kon$ as the set of minima of
the functionals \erf{mors}.

Finally we note that
if the background is reducible, then the Morse functionals $\F$ possess
a systematic degeneracy. Namely, if $h$ is an element of the stabilizer
$\cals_{\!\abar}^{}$ of the background \bara, then we have
   \be  \F[gh] = |A\gt{gh} - \abar|^2 = |A\gt{gh} - \abar\gt h|^2
   = |(\ag - \abar)\gt h|^2 = |(\ag - \abar)|^2 =\F[g] \, . \labl{eig}
Here one uses the fact that the difference of two connections transforms
homogeneously, and that the norm $|\cdot|$ is invariant under the action
of the gauge group. Conversely, if for a given
$h \in \calg$ \erf{eig} holds for all $A \in \cala$ and all $g\in\calg$,
then $h$ is in fact an element of the stabilizer of \bara.

\nsection{Fundamental modular domains}

In general the subset $\kon=\kon_{\bar A}$ of \cala\ contains
absolute minima as well as relative minima of the functionals $\F$. The subset
  \be  \L\equiv\LA:= \left\{ A \in \cala \mid \F(g) \geq \F(\bfe)
  \ \mbox{for all}\ g \in \calg \right\} \subseteq\kon_{\abar}^{}  \ee
of {\em absolute\/} minima of $\F$ contains at least one representative of
any gauge orbit \calo\ \cite{sefr,dezw3} and for topologically simple
\st s is properly contained in \kon\ \cite{dezw2}.

Using again repeatedly the fact that the difference of two connections
transforms homogeneously, and that the inner product \erf{skp} is
invariant under gauge transformations, we can deduce that
  \be \begin{array}{l}
  |(\xi B + (1-\xi )C)\hg-\abar|^2 - |(\xi B + (1-\xi )C)-\abar|^2 \\[1.7 mm]
  \hsp5
  = |\xi(B-C)\hg+(C\hg-\abar)|^2 - |\xi(B-C)+(C-\abar)|^2 \\[1.7 mm] \hsp5
  = 2\xi\,((B-C)\hg,C\hg-\abar)-2\xi\,(B-C,C-\abar)+|C\hg-\abar|^2-|C-\abar|^2
  \\[1.7 mm] \hsp5 = 2\xi\,(B-C-B\hg+C\hg,\abar)+|C\hg-\abar|^2-|C-\abar|^2
  \\[1.7 mm] \hsp5 = \xi\, ( |B\hg-\abar|^2 - |B-\abar|^2 ) + (1-\xi )\,
  ( |C\hg-\abar|^2 - |C-\abar|^2 )  \end{array} \labl:
for any $\abar,B,C\in\cala$ and for any $\xi\in\reals$.
As a consequence, if both $B$ and $C$ belong to \LA\ so that both
$|B\hg-\abar|^2 - |B-\abar|^2$ and $|C\hg-\abar|^2 - |C-\abar|^2$ are
positive, then for all $\xi\in[0,1]$ the left hand side of \erf: is positive
as well, and hence the connection $\xi B+(1-\xi)C$ belongs to \LA, too.
This shows that, just as \konA, the subset \LA\ of \cala\ is convex.
Furthermore, as \LA\ is contained in the bounded set \konA, it is bounded,
too. (However, just as e.g.\ the \infdim\ unit sphere, \LA\ is not compact.)

Now as \L\ is a convex subset of an affine space and hence topologically
trivial, the topological non-triviality of \calm\ must stem from the
fact that upon projection onto \calm\ some points of \L\ must be identified
in a non-trivial manner. A priori such identifications may take place
for boundary points of \L\ as well as in the interior of \L.
We will now show that the identification in the interior precisely
amounts to dividing out the stabilizer \calsab.

For reducible backgrounds the systematic degeneracy \erf{eig} of the
functionals $\F$ implies that in particular their absolute minima are
degenerate. Thus in order to obtain a \fmd\ one must divide out at least
the action of the stabilizer of the background from \L. Actually, the
latter procedure is already sufficient to obtain a modular domain. To see this,
define \LO\ as the subset of \L\ that consists of all connections
in $\Lambda$ for which the only gauge copies contained in $\Lambda$ are
precisely those related by elements of the stabilizer of the background.
Owing to \erf{alt} one has the identity
  \be  \FAA[g] = |\bar A^g - \bara|^2 = |g^{-1} \delo g|^2, \labl{faa}
which implies that on the orbit of \bara\ the minimum of the Morse
functional is zero. Any gauge transformation $g$ for which $\FAA[g]$ has
this minimal value is covariantly constant \wrt $\bara$ and thus in
the stabilizer \calsab\ of the background. This shows that \LO\
contains at least the background $\bar A$ and is thus not empty. In addition,
the difference $\L\setminus \LO$ is contained in the boundary of \L,
  \be  \Lambda \setminus \LO \subseteq \partial \Lambda \,.  \labl[
Namely (compare \cite{vanb3}), let $C$ be a connection in $\L\setminus \LO$.
When taking $B=\bar A$ in \erf:, the first term on the right hand side
is equal to $\xi\, |\bara^g - \bara|^2$ and is thus strictly positive
for any $\xi> 0$ and all $g \in \calg \setminus \calsab$, while
the term proportional to $1-\xi$ is non-negative for all $\xi\in[0,1]$
since $C$ corresponds to a minimum of $\FAC$. As a consequence, any point
on the straight line between $\bar A$ and $C$ is an absolute minimum
and is in fact contained in \LO. Only for $\xi=0$, i.e.\ at the connection
$C$ itself, the minimum has further degeneracies. We conclude that
$\L\setminus \LO$ cannot have inner points, since otherwise the straight
line between $C$ and \bara\ had to contain
minima with larger degeneracy for some $\xi\in(0,1)$. This proves \erf[.

Furthermore, the stabilizer of any connection in the interior of \LA\ has
to be contained, as a subgroup of \calg, in the stabilizer of the background.
This holds because the interior of \L\ is contained in $\LO$, and because
by definition in $\LO$ the degeneracy of the
Morse functional is trivial, namely equal to the stabilizer of the background
$\bara$. Now if $g$ is in the stabilizer of $A$, then
  \be \F[gh] = |A\gt{gh}-\bara|^2= |A\gt{h}-\bara|^2= \F[h]  \ee
for all $h\in\calg$.
Thus if the stabilizer of $A$ is not contained in the stabilizer of \bara,
then there is an additional degeneracy, and hence $A\in\Lambda \setminus
\LO \subseteq \partial \Lambda$. In particular, if $A\in\L$ and \bara\
belong to the same stratum, then their stabilizers
are in fact identical rather than only conjugated subgroups of \calg.
Given the partial ordering of strata induced by the inclusions of conjugacy
classes of the stabilizers, our result implies that the interior of $\Lambda$
contains only elements of strata with equal or `less' symmetry as the
background. Specializing to the case of irreducible background, it follows
that reducible connections are necessarily on the boundary of \L.

With the above result it is easy to give locally a more precise
description of the pre-images of the strata of \calm\ in $\Lambda$.
Suppose we are dealing with a connection \bara\ as the background which
has non-trivial stabilizer \calsab.
Let $U$ be a neighbourhood of \bara\ that  is contained
in the interior of $\Lambda$, and $\tilde U$ the intersection of the stratum
of \bara\ with $U$. Since all elements $A$ in $\tilde U$ have identical
stabilizer, any connection
  \be \xi \bara + (1-\xi) A, \qquad \xi \in [0,1] \,, \ee
has the same stabilizer as well, and is hence contained in $\tilde U$. This
shows that $\tilde U$ is the intersection of a linear subspace of $\PSI$ with
$U$. (This description applies only locally, and in general the same stratum
may also have additional points, with stabilizer conjugate but not identical
to \calsab, on the boundary $\partial \Lambda$. We will see in section
\ref{secsu2} that this is in fact the case for the U(1) stratum of a SU(2)
gauge theory over $S^4$.)

We are now in a position to specify the true fundamental modular domains:
for any background \bara, they are given by $\LA/\cals_{\!\abar}^{}$.
For the objects that are obtained from these \fmd s by boundary identification,
such that they are isomorphic to the configuration space \calm,
we will use the notation \La:
  \be  \La\equiv\La_{\abar}= \LA\restr{\rm bound.id.}/ \cals_{\!\abar}^{}
  \,, \ee
which reduces to
  \be  \La= \L\restr{\rm bound.id.}/Z(G) = \L\restr{\rm bound.id.} \ee
in the case of an irreducible background \bara. After taking into account
the boundary identifications in this manner, \La\ can be shown to be
paracompact \cite{asmi2}. As $Z(G)$ leaves all connections fixed,
in the sequel we will no longer mention the presence of $Z(G)$ explicitly.
Note that as soon as the stabilizer of the background is non-trivial,
the action we have to divide out is
non-trivial as well, and in fact it possesses at least one fixed point, namely
the background itself. When fixing the gauge around a reducible connection
$\bar B$ (e.g.\ $\bar B=\anull$), this property ensures that $\bar B$
which is a \smi point of $\L_{\bar B}$ becomes a singular point of
$\L_{\bar B}/\cals_{\!\bar B}$, and in fact a singular point of $\La_{\bar B}$.

Let us for the moment assume now that the background \bara\ is irreducible.
Then in the transition to \La\ in fact any irreducible connection on the
boundary $\partial\LA$ of \LA\ has to be identified with other irreducible
connections that lie on $\partial\LA$.
This can be seen as follows. Assume that $B\in\partial\LA$ is
irreducible. Then instead of \LA\ we can alternatively consider the
corresponding subset $\L_B\subset\cala$ that is obtained by taking $B$ as a
background. Now the same argument that was used in the proof of \erf[
shows that $B$ is a \smi point of $\L_B$. In addition, as just mentioned,
upon projection to \calm\ inner points do not get identified and hence are
projected to \smi points of \calm. But the only way in which
the connection $B$, now considered again as a point of $\partial\LA
\subset\LA$, can project to a \smi point of \calm\ is that a
neighbourhood of $B$ on the boundary (which can be taken to consist only of
irreducible connections because the reducible connections are nowhere dense)
has to be identified with a different neighbourhood on the boundary of \LA.
(The latter neighbourhood has to consist of boundary points,
because inner points do not have non-trivial copies in \LA.)
In contrast, if $B$ is reducible, then the identifications of boundary
points have to take place in such a manner that the point remains singular.
Roughly speaking, there are `less' identifications for reducible
than for irreducible connections.

The previous results imply in particular
that \La\ does not have any boundary points except for reducible connections.
The latter are, strictly speaking, no boundary points either, because
by the usual definition a boundary of a manifold is another manifold of
codimension one that is patched to the manifold in a specific manner; in
the case of singular connections the codimension
is in general infinite, and the patching is much more complicated.

It is worth emphasizing that even if the subset of reducible connections
is `small', in the context of quantum field theory its presence
cannot simply be ignored. (For instance, Green functions
are distributions, and hence one has to analyze carefully
whether a set of measure zero can be disregarded.)
Unfortunately, only little is known about
possible effects on quantum physics that are due to the presence of
singularities in the configuration space. Indeed we expect that the
reducible connections have to be treated with still more care.
For instance, it might be necessary to resolve the singularities,
in a consistent manner analogous to the blowing up of orbifold
singularities of \findim\ complex \cite{GRha} or symplectic \cite{AUdi,garo}
manifolds. This would induce an additional source of non-triviality for
the topology of \calm, and thus also for the anomaly structure
if the theory is coupled to chiral fermions.
Now in a Hamiltonian formulation it can be shown \cite{armm} 
 that classical trajectories are always contained in one fixed stratum of the
configuration space, so that when dealing with the classical field theory we
can restrict ourselves to a fixed stratum, which is in itself a smooth
infinite-dimensional manifold. Therefore we expect any influence
on the main stratum arising from the
singularities of \calm\ to be a genuine {\em quantum\/} effect.

Finally we recall from the introduction that by the choice of
a specific (isomorphism class of) principal
bundle $P$ we have fixed the instanton number. Thus when allowing for
arbitrary instanton number $k$, i.e.\ considering the \YM\ theory with
arbitrary asymptotic conditions at infinity, the fundamental modular
domain is in fact the disjoint sum over $k\in\zet$ of the modular domains
for each value of $k$. (The collection of the orbit spaces
for all $k\in\zet$ is sometimes referred to as the extended
orbit space \cite{asmi2}.)

\nsection{The pointed gauge group}

In this section we investigate the relation between the region
$\Lanul:=\Lnull\restr{\rm bound.id.}$ of absolute minima of the Morse
functionals $\FAnull$ for the background \anull, and another object, namely
the orbit space  $\cala/\calgo$ \wrt the so-called {\em pointed gauge group\/}
\calgo. The latter group plays an important r\^ole for detailed investigations
of the Riemannian geometry of the configuration space
\cite{atjo,naRa,mivi,koro,sinG2}; it is defined as
  \be  \calgo:= \{ g \in \calg \mid g(x_0) = \one \}    , \ee
with $x_0 \in M$ an arbitrary, but fixed, \st\ point.
Suppose we have performed all boundary identifications in \Lnull, thereby
obtaining an infinite-dimensional variety \Lanul, which in contrast to
\Lnull\ is not embedded in an affine space. We claim that
  \be  \Lanul \cong \cala / \calgo  \,, \labl;
i.e.\ that \Lanul\ is isomorphic, as a manifold endowed with a $G$-action,
to the space \cala\ of all connections divided by the pointed gauge group.
The fact that both objects are in particular diffeomorphic shows that
\Lanul\ is not only a variety, but even a smooth \infdim\ manifold.

To prove \erf;, we first note that a set of representatives for
  \be \calg / \calgo \cong  G  \ee
is given by the constant gauge transformations, i.e.\ by the stabilizer
of the background $\anull$. This is simply
because any gauge transformation $g\in\calg$ can be written as
  \be g = \gc\, (\gc^{-1} g) \,  ,\ee
where $\gc$ denotes the constant gauge transformation with
value $\gc(x)=g(x_0)\in$\,SU(2) independent of $x$. Also, the only
constant gauge transformation that is an element of \calgo\ is the unit
element \bfe.  We now map any element of \Lanul\ (or more precisely,
its pre-image in \Lnull) on its equivalence class modulo \calgo.
We have to prove that this map is injective and surjective. To
show that the map is injective, suppose that $A,B \in \Lanul$ are mapped on
the same element of $\cala/\calgo$. Then $A$ and $B$ are in the same
equivalence class \wrt \calgo, i.e. $A =B\hg$ with $g$ an element of
\calgo. On the other hand, the only gauge copies left in \Lanul\
are those related by constant gauge transformations; therefore
$g$ must be the identity, and hence $A = B$, which proves injectivity.
To show that the map is surjective we start with an arbitrary connection
$A \in \cala$. Since the orbit of any connection has a representative in
\Lnull\ (and hence in \Lanul) we can find an element $B\in\Lanul$ and an
element $g\in\calg$ such that $B\hg=A$. Now decompose $g$ as $g =\gc\go$ with
$\go\in\calgo$ and $\gc$ a constant gauge transformation. Then
  \be  B\gt{\gc}_{}=A\gt{\gome}_{} \,. \labl{oc}
Now due to $\gc\in\calsnull$ the left hand side of \erf{oc} is also in
\Lanul, while the right hand side is in the same equivalence class
modulo \calgo\ as $A$. Thus for each element of $\cala/\calgo$ there exists
an element of \Lanul\ that gets mapped to it, which proves surjectivity.

Furthermore, the spaces \Lanul\ and $\cala/\calgo$ both
carry a group action of the finite-dimensional structure group $G$.
In the case of \Lanul\ this action is defined by applying a constant
gauge transformation $\gc$ on $A \in \Lanul$\,; this is well-defined because
$A\gt{\gc}$ is an element of \Lanul, too. For $\cala/\calgo$ the $G$-action
is again defined by use of constant gauge transformations: the orbit
$\goorb A$ containing $A$ is mapped to the orbit $\goorb{A\gt{\gc}}$
of $A\gt{\gc}$, in other words $(\goorb A)\gt{\gc}:= \goorb {A\gt{\gc}}$.
This is well-defined because $\calgo$ is a normal subgroup:
choosing a different representative $A\hg\in\goorb A$ with $g \in \calgo$,
the fact that \calgo\ is normal in \calg\ means that there is a $g'\in\calgo$
such that $ g \gc = \gc g'$, and hence $(A\hg)\gt{\gc}= (A\gt{\gc})\gt{g'}
\in \goorb {A\gt{\gc}}$.
The fact that for any $A \in \Lanul$, $A\gt{\gc}$ is mapped under the
isomorphism on $\goorb {A\gt{\gc}} =(\goorb A)\gt{\gc}$ shows that
both group actions coincide, or, more precisely, that the isomorphism
intertwines the group actions.

Finally, let us note that analogous arguments as given above for the
background \anull\ show that a similar relation as \erf; holds
whenever the background is a pure gauge. Namely, if
  \be  \abar= \anull^{\,\gbar}=\gbar^{-1}\,{\rm d}\gbar \,, \ee
then
  \be  \LA\restr{\rm bound.id.} \cong \cala / \calgog  \,, \ee
with
  \be  \calgog:=\{g\in\calg\mid (\gbar g \gbar^{-1})(x_0)=\one\}
               = \gbar^{-1} \calgo\, \gbar \,, \ee
and the action of the structure group is via the stabilizer $\calsab=\bar g
G \gbar^{-1}$ rather than via the constant gauge transformations.

\nsection{Geodesic convexity}\label{secgc}

We are now going to discuss the consequences of the convexity of \LA\
for the configuration space. First note that, in contrast to \LA,
due to the division by \calsab, the \fmd\
\LG\ is in general not a subset of an affine space, and hence we
do not have the notion of convexity any more. However, we can show
that the main stratum in \LG\ is still {\em geodesically convex}, i.e.\
any two non-singular points in \LG\ can be joined by a geodesic which
only contains non-singular points of \LG.
(Of course, the fact that a space is geodesically convex
does not mean at all that it is topologically trivial \cite{atjo}.)
This can be seen as follows.

Suppose that $\pb,\,\pc$ are two non-singular points in
\LG, and let $B$ be a representative of $\pb$ in \LA.
Then consider $B$ instead of \bara\ as the background connection,
and let $C$ be a representative of $\pc$ in \LB.
It follows \cite{bavi} that any geodesic through $\pb$ in the orbit space,
and hence in particular any geodesic joining $\pb$ with $\pc$,
is given by the projection of a straight line through $B$ in \cala.
Now since \LB\ is convex, the straight line connecting $B$ and $C$ is contained
in \LB, and it projects down to a geodesic in \LG.
Furthermore, no reducible connection is contained in the straight line
that connects $B$ with $C$ so that the corresponding geodesic does not
hit a singularity. This is because, as seen above, for irreducible
background $B$ any reducible connection in \LB\ lies
on the boundary $\partial\LB$.
Note that this is still true if $C$ is a point on the boundary of \LB\
(the straight line from $B$ to $C$ then meets $\partial\LB$, but only
in the single point $C$ which by assumption is irreducible).
This implies that the result still holds after boundary identifications
are taken into account, i.e.\ for \LGb.

It is natural to ask whether strata other than the main stratum are
geodesically convex as well, and one may try to investigate this
problem along similar lines as above. Inspection shows, however, that
for non-main strata the situation is generically much more complicated
than above. At a technical level,
the main obstacle is that one cannot easily verify whether
to any two points in the stratum one can find representatives in a
\fmd\ for which the stabilizers are identical rather than only
conjugate subgroups of \calg.

To conclude, let us comment on the physical meaning of geodesic convexity.
When interpreting our result in a Hamiltonian picture, it
provides information about the obstructions that the
geometry of the configuration space imposes on the classical motion.
Notice that in pure \YM\ theory classical trajectories can be
characterized by their instanton number which is a topological quantity,
so that requiring the instanton number to have a fixed value amounts to a
{\em kinematical\/} restriction to the motion. There is also a
{\em dynamical\/} obstruction stemming from the fact that classical
trajectories are contained in a stratum of fixed stabilizer type.
Our result shows that within the main stratum there is no further
obstruction of the latter type. The only possible further restrictions on
the motion must then stem from the fact that the geodesics are {\em not\/} the
classical trajectories, because in the Hamiltonian approach also a
potential must be taken into account \cite{bavi}.
In the case of the \YM\ action \erf{lagr}, the relevant potential is
$ V=\frac1{4{\rm g}^2} \ixxx \tr F_{ij}F^{ij}$, where integration
is over some time slice $\cal C$.

\nsection{Singularity structure of the SU(2) configuration space}
\label{secsu2}

\subsection{Reducible connections}

As an application, we now consider in detail the case $G=\,$SU(2).
As already mentioned in section 3, the holonomy group $H$ is a Lie
subgroup of the structure group $G$. For $G=\,$SU(2), this requirement
leaves the following possibilities. First, the holonomy group $H$
can be SU(2) or SO(3). Then the centralizer of $H$ is just
the center $\zet_2=\{\pm\one\}$ of SU(2). Thus the
stabilizer is trivial and hence the connection is irreducible.
Secondly, the holonomy group may be $H=\,$U(1). Then the stabilizer is
also isomorphic to U(1) as an abstract Lie group, and the analysis is less
trivial than in the previous case. Finally, the holonomy group may be
trivial so that the stabilizer is isomorphic to SU(2); as we will see,
this situation is best described as a special case of the connections
with stabilizer isomorphic to U(1).

To enter the discussion of connections with trivial or
U(1) holonomy, we recall that we assume the \st\ manifold $M$ to be $S^4$,
so that in particular its second cohomology group vanishes, $H^2(M,\zet)=0$.
It can be shown \cite{sinG} that for $G=\,$SU(2) the vanishing of
$H^2(M,\zet)$ implies that the instanton number of any reducible
connection vanishes. Namely, let $V \cong
\complex^2$ be a \twodim\ complex vector space carrying
the defining representation of $G=$ SU(2). If the connection has only U(1)
holonomy, the associated vector bundle $E:= P \times_{\rm SU(2)}V$
splits into the direct sum of two complex line bundles over $M$,
  \be  E = P\times_{\rm SU(2)} V_1 \,\oplus\, P\times_{\rm SU(2)} V_2 \,. \ee
As in our case the second real cohomology of the \st\ $M$ is trivial,
both line bundles and hence also $E$ and $P$ have to be trivial,
i.e.\ $P=M\times G$. In particular, gauge transformations $g\in\calg$ can
be considered as functions from $M$ to the structure group $G$.

It must also be noted that the space of maps from $S^4$ to SU(2) is
topologically non-trivial, $\pi_4({\rm SU(2)}) = \zet_2$. As a
consequence, for structure group SU(2) the gauge group \calg\ is {\em not\/}
connected. However, there are no fixed points \wrt the non-identity component
of \calg. This holds because any stabilizer is isomorphic to a Lie subgroup
of SU(2), and all these subgroups are connected.

\subsection{The U(1) stratum}

Assume now that $A\in\cala$ is a connection with nontrivial stabilizer,
i.e.\ that the stabilizer $\calsa$ contains at least a U(1) subgroup of \calg.
This U(1) subgroup of \calg\ in general does not consist of constant
gauge transformations. However, we are only interested in gauge orbits,
and the stabilizers of different points on an orbit are related via
conjugation by gauge group elements. We will
now show that we can always find a representative of the orbit
for which the stabilizer U(1) does consist of constant gauge transformations.
We first notice that because of the  direct product structure $P = M \times G$,
we can write down all quantities of interest
in terms of functions over \st. In particular, we can write a
covariantly constant section $\s$ in the adjoint bundle
$P \times_{\rm SU(2)}^{}{\rm su(2)} \cong M \times {\rm su(2)} $
$=S^4\times {\rm su(2)} $ as
$\s(x) =\sum_{a=1}^3 \s_a(x) \tau^a$, where $\{\tau^a\mid a=1,2,3\}$ is
a basis of the Lie algebra su(2). The fact that $\s$ is covariantly
constant implies that the `length squared' $\,\tr(\s(x)^\dagger\s(x))$ of
$\s(x)$ is in fact independent of $x\in S^4$.
Further, $G$ acts transitively on elements $\tau$ of the same length; this
suggests to relate \s\ to some fixed arbitrary element $\tauo$ of su(2)
that has the same length as \s. Thus we set
  \be  \s(x) = g(x)\, \tauo\, g^{-1}(x) \,. \labl{sg}
Now $G$ does not act freely on elements of equal length; rather, the
stabilizer of $\tauo$ is the subgroup $H$ that is spanned by all
elements of the form $\eE^{\ii\tauo\phi(x)}$ for some function $\phi$.
As a consequence, for any
$x \in M$ the equation \erf{sg} does not determine uniquely an element
of $G$, but rather specifies one coset of $H$ in $G$. This coset is isomorphic
to the circle $S^1$, and hence we are dealing with a bundle over $S^4$
with fiber $S^1$. Now any 1-sphere bundle over $S^4$ is a Cartesian
product $S^4\times S^1$ \cite[section 26.5]{STee}. This implies that
we can find a global section $g(x)$ in this bundle, which can be interpreted
as an element of \calg. Thus \erf{sg} provides us with a well-defined gauge
transformation $g\in\calg$. (Note that this is in contrast to
the situation in two dimensions \cite{hkprs2}, where there is a different
bundle corresponding to each integer \cite[section 26.2]{STee}.)

Applying the gauge transformation $g(x)$ that is determined in this
manner by the section \s, the fact that \s\ is covariantly constant,
$\nabla^{}_{\!\!\!A_{}} \s = 0$, yields
  \be 0 = \nabla^{}_{\!\!\!A_{}^{g^{-1}}} \s^{g^{-1}}
 = \nabla^{}_{\!\!\!A_{}^{g^{-1}}} \tauo = [ A\gt{g^{-1}},\tauo] .  \ee
This implies that we can always find one gauge equivalent
representative for a reducible connection that is of the form
  \be  \amu(x)=\amutau(x):= \ii\tauo\,\fmu(x) \,. \labl{abel}
The field strength corresponding to this connection
is $ F_{\mu \nu}(x) =\ii\tauo\,f_{\mu \nu}
(x)$, where $ f_{\mu\nu} := \partial_\mu \fnu - \partial_\nu \fmu $.
In \erf{abel}, the quantities $\fmu$ are arbitrary (sufficiently smooth)
real-valued functions $\fmu(x)$ on $S^4$; in the sequel we assume that a
fixed choice for these functions has been made.
It is easy to verify that the gauge configuration \erf{abel}
is invariant at least under the U(1)-subgroup of constant gauge
transformations $\{\eE^{\ii\theta\tau}\mid\theta\in[0,2\pi)\}\subset\,$SU(2)
that is generated by $\tauo$.  As promised, the stabilizer U(1) of
the representative \erf{abel} of the gauge orbit consists of
constant gauge transformations. The fact that we can always find such a
representative means that the orbits of connections with stabilizer
isomorphic to U(1) form one single stratum of the configuration space \calm.

Let us consider the configurations \erf{abel} more carefully and investigate
at which point(s) the background gauge condition
 $\delos(A\hg-\abar)=0$ is satisfied.  We have
 $\amu\hg=\gg\amu\, g+\gg\dmu g=\ii\gg\tauo g\,\fmu +\gg\dmu g$, and hence
  \be  \begin{array}{l} \dmU\amu\hg=\ii\dmU(\gg\tauo g)\,\fmu+\ii\gg\tauo g\,
  \dmU\fmu +\dmU(\gg\dmu g) \\[2 mm] \hsp{2.3}
  = \ii\gg\tauo g\,\dmU\fmu + \ii(-\gg\dmugg\tauo g +\gg\tauo\dmu g)\,\fmu
  -\gg\dmUgg\dmu g +\gg\dmu\dmU g. \end{array}\labl4
In the following we take the background to be the reducible connection
$\abar=\anull$. (The results will be independent of the choice of
background, but explicit calculations are much more involved for any
other background.) Thus the background gauge requirement is $\dmU\amu\hg
=0$; with \erf4, this reads $\,\ii\tauo\,
\dmU\fmu=\ii(\dmugg\tauo-\tauo\dmugg)\,\fmu +\dmUgg\dmugg-\dmu\dmUgg $, i.e.
  \be  \ii\tauo\,\dmU\fmu= \ii[\dmugg,\tauo]\,\fmu-\dmU(\dmugg) \,. \labl1
To determine the general solution to this second order \difeq\ for $g$
would be a difficult task. However, we can find a particular solution
by simply recalling the analogous problem in electrodynamics; thus we
make the ansatz that $g$ is of the special form
  \be  g(x)=\exp(\ii\tauo\gamma(x)) \labl5
(thus in particular $g$ is connected to the
identity). Then $\dmugg=\ii\tauo\,\dmu\gamma$, so that
  \be  \amu\hg=\ii\tauo\,(\fmu+\dmu\gamma) \,. \labl{Expl}
Also, (\ref1) becomes $\tauo\,\dmU\fmu= -\tauo\,\dmU\dmu\gamma$;
we thus end up with the differential equation
  \be  \partial^2\gamma=-\dmU\fmu \labl{dd}
for the function $\gamma$. The general solution of \erf{dd} is
  $    \gamma(x)=\int_{M}{\rm d}^4y\, H(x,y)\, \dmU\fmu(y) + c \,, $
where $H$ denotes the Green function of the Laplacian
$\partial^2$ on $M$, i.e. $\partial^2_x H(x,y)=-\delta(x-y)$
with $\delta$ the delta function on $M$;
for $S^4$, it reads $H(x,y)=\onehalf\, |x-y|^{-2}/{\rm Vol}(S^3)$.
The constant contribution $c$ to $\gamma(x)$ is the general solution of the
homogeneous equation $\partial^2\gamma=0$, as follows from the fact that
our \st\ manifold is compact and without boundary.

For completeness we note that in fact $g$ as defined by \erf5 and \erf{dd}
is an element of the relevant Sobolev gauge group $\calg^k$ with norm
$  |g\restr k = \sum_{l=0}^k\, \int_M {\rm d}^4 x\,
  \tr( (D^l g)^*(D^l g) ) \,. $
Here $D^l$ is a shorthand for some multiple derivative of order $l$
(also, we assume that $k > \dim(M)/2 $ so that we can use the Sobolev
inequality, compare \cite{mivi}).
Standard regularity results for the Poisson equation \erf{dd} immediately
give estimates on a suitable Sobolev norm of $\gamma$. This can in turn
be used to get similar estimates for the norm of $g$.
(For instance, for $l=0$ we have to integrate\, $\tr(g^* g)= \tr \one = 2$
over the compact \st\ $M$, which gives a finite result.
Furthermore, since any derivative of $g$ can be written as a
linear combination of unitary matrices with coefficients being
polynomials in the derivatives of $\gamma$, we can control the norms
of the derivatives of $g$, too.)

Thus we have shown that on any gauge orbit of reducible connections
we can identify a
connection which both lies in the gauge slice $\PSI$ and is of the
specific form \erf{abel}. Of course, not all of these connections
will also belong to the \fmd\ $\Lnull/SU(2)$
which is only a subset of the solutions to the gauge condition. To decide
which of the above solutions belongs in fact to the modular domain,
we would have to look for the absolute minima of
the functionals \erf{mors} for $\abar=\anull$, i.e.\ of $\F[g] =|\ag|^2$.
For $A=\atau$ \erf{abel} and $g=\eE^{\ii\tau\gamma}$ \erf5, one has
  \be  \F[g_{\rm min}] ={\rm tr}(\tauo^2)\,
  \ixxxxm \left\{ (\fmu+\dmu\gamma)(\fmU+\dmU\gamma) \right \} .\labl{i1}
Upon partial integration and use of the differential equation \erf{dd}
we can rewrite \erf{i1} as $\F[g_{\rm min}]=\F[\bfe] - {\rm tr}(\tauo^2)\,
\ixxxxM \dmu\gamma\, \dmU\gamma$, or equivalently as
  \be  \F[g_{\rm min}] =  \F[\bfe] + {\rm tr}
  (\tauo^2)\, {\displaystyle\int}\!\!\ixxxxm \,{\rm d}^4 y\; \dmU a_\mu(x)
  \,H(x,y)\, \partial^\nu a_\nu(y) \,,  \ee
i.e.\ in the form of a self-energy. Obviously, for generic functions $a_\mu$
there is no simple way to tell whether this value is the absolute minimum of
$\F$ or not. Hence in spite of the fact that we can identify on each singular
gauge orbit a connection that belongs to the gauge slice, it would be very
difficult to determine the U(1) stratum of the configuration space via the
analysis presented above. Fortunately we can bypass this problem by employing
the relation with electrodynamics in a different manner; this will be
described in subsection \ref{secsu2}.4. However, before coming to that, in the
next subsection we will deal with a particularly simple situation where the
above analysis can indeed be applied.

\subsection{The stratum with SU(2) stabilizer}

Namely, let us consider the case where the holonomy group is $H=\{\one\}$;
then the stabilizer is isomorphic to SU(2). Thus in particular the
stabilizer contains a U(1) subgroup, and therefore we can use the result
of the previous subsection to conclude that the gauge slice contains
representatives of the form \erf{abel}. In addition, however, the fact
that the holonomy vanishes implies that the field strength
$\ii \tau f_{\mu\nu}$ for such connections vanishes, i.e.\ that the
connection is a pure gauge. In this case we may integrate \erf{dd} to
  \be  \fmu+\dmu\gamma=\cmu={\rm const} \,; \labl{soco}
conversely, this solution exists only if $f_{\mu\nu}(x)\equiv0$. Inserting
the result \erf{soco} into the formula for the gauge transform of $A$ yields
  \be  \amu\hg=\tauo\,(\fmu+\dmu\gamma)=\tauo\,\cmu. \labl{expl}
This means that, while any constant \gtrafo\ out of the subgroup corresponding
to $\tauo$ leaves $A$ invariant, there is an analogous non-constant
\gtrafo\ such that $A\hg$ is a {\em constant\/} multiple of $\tauo$.
Further, the derivation of the result does not determine the constants
$c_\mu$; thus {\em any\/} choice of constant $\amu=\cmu\tauo$ satisfies the
gauge condition (\ref4). In particular, $\cmu=0$ yields a solution,
i.e.\ $\amu(x)\equiv0$ belongs to the orbit of $\amutau$.
Thus any configuration of the type \erf{abel} with vanishing field
strength is in fact gauge equivalent to the \confi\ $\anull$.
We also note that any value of the constants
$\cmu$ leads to a connection on the orbit of \anull\ that lies in the
region $\PSI$. In contrast, there is only one point
on this orbit which is also contained in $\Lambda$, namely just
the connection $\anull$ itself. Namely, combining \erf{soco} and
\erf{i1}, we see that we have to set all constants $\cmu$
equal to zero, i.e.\ $(\atau)\hg=\anull$, in order for $|(\atau)\hg|^2$
to be an absolute minimum. (Clearly, this absolute minimum is at zero, which
is just a special case of the analogous statement for the functional
\erf{faa}.) It is also clear from the
general arguments of section 4 that the gauge transformations which
preserve the absolute minimum are precisely the constant ones.)

In short, we have shown that the only connections with maximal stabilizer
in a SU(2) \YM\ theory on the \st\ $S^4$ are the pure gauges, and that after
fixing the gauge around the \confi\ $\anull$, the only such connection
contained in \Lanull\ is $\anull$ itself.

We can also give explicitly the specific non-constant gauge transformations
that combine with the U(1) to form a SU(2) which leaves the original
connection invariant. Namely, the stabilizer is simply the conjugate of
the stabilizer of $\anull$ by $\eE^{\gamma(x)\tau}\in\calg$, with
$\gamma(x)$ a solution of \erf{soco} with $\cmu= 0$; its elements
are of the form $\eE^{\gamma(x)\tau} g\, \eE^{-\gamma(x)\tau} $, where $g$
is an arbitrary constant gauge transformation.

\subsection{The reducible strata as a $\zet_2$-orbifold}

We now describe the set of all orbits with non-trivial
stabilizer from another point of view. Our starting observation is that
the equations \erf{abel} and
\erf{Expl} suggest that this set is related to the configuration space
of a U(1) gauge theory, i.e.\ of electrodynamics. Indeed, we can construct a
map from the orbits of vector potentials $a_\mu$ of a U(1) gauge theory on the
orbits of reducible connections of the SU(2) theory by mapping the orbit
containing $a_\mu$ on the orbit containing $\ii a_\mu \tau$. This is
well-defined because \erf{Expl} ensures that if $a_\mu$ and $b_\mu$ are
on the same orbit, i.e. $b_\mu=a_\mu+\partial_\mu\gamma$, then also the images
of $a_\mu$ and $b_\mu$ are related by a gauge transformation, namely by \erf5.

In a second step we analyze which orbits of the U(1) theory are mapped on the
same orbit of the SU(2) theory. Thus assume that $A_\mu = \ii \tau a_\mu$ is
related to $B_\mu = \ii \tau b_\mu$ by some element $g$ in the gauge group of
the SU(2) theory (that is not necessarily of the form \erf5). Then in
particular the associated field strength tensors are related by
  \be F^{\mu \nu}_{(A)}(x) = \ii \tau f^{\mu\nu}_{(a)}(x) =
  g^{-1}(x)  F^{\mu\nu}_{(B)}(x) g(x) = \ii g^{-1}(x)\tau g(x) f^{\mu\nu}
  _{(b)}(x) \,.  \ee
Without loss of generality let us choose $\tau\in\,$su(2) to be the Pauli
matrix \taudr; then for any fixed $x\in M$ we have to look for all elements $g$
of SU(2) for which $g^{-1} \taudr g$ is proportional to \taudr.
For this it is necessary that either $g$ is an element of the U(1) group
generated by \taudr, i.e. $g=\exp(\ii\theta\tau^3)$, or that $g$ is of the form
  \be g =  \left ( \begin{array}{cc} 0  & \eE^{\ii \theta} \\
                     -\eE^{-\ii \theta} & 0   \end{array} \right )
                 = \ii \tau^2  \eE^{\ii \theta \tau^3} .  \labl{neu5}
Thus $g$ must be an element of the following subgroup $H$ of SU(2):
topologically, $H$ is the disjoint union of two circles and hence not
connected. Algebraically, $H$ contains a normal U(1) subgroup spanned by
all elements of the form $\exp(\ii\theta\tau^3)$; the two cosets relative to
that subgroup have representatives $\one$ and $\ii \tau^2$ and form thus a
$\zet_2$ subgroup; hence $H$ is a semi-direct product
  \be  H=\zet_2 \semitimes \mbox{\,U(1)} \,.  \ee

If a gauge transformation $g\in\calg$ connects gauge potentials of the
form \erf{abel}, then for any $x \in M$ it must lie in $H$, i.e.\ $g(x)$
is of the form described above, with the real number $\theta$ replaced
by some real function $\gamma(x)$. Now our \st\ $M=S^4$ is connected,
and the elements of the gauge \calg\ correspond to {\em continuous} mappings
from $M$ to SU(2). Therefore the value of $g(x)$ lies in one and the same
component of $H$ for all \st\ points. If it is in the component connected
to the identity, then the gauge transformed connection is as already obtained
in equation \erf{Expl}. In the other case, i.e.\ for $g$ of the form
\erf{neu5} with $\theta$ replaced by $\gamma(x)$, we get instead
  \be A_\mu^g = -\ii \tau(a_\mu + \dmu \gamma) \,. \ee
This implies that the orbits of the U(1) gauge theory that correspond to the
functions $a_\mu$ and $-a_\mu$ are mapped on one and the same orbit of the
SU(2) theory. We conclude that the set of orbits of trivial or U(1) holonomy
is a $\zet_2$-orbifold of the configuration space of electrodynamics. The only
singular point of this orbifold is given by the single orbit with trivial
holonomy which corresponds to $a_\mu \equiv 0$, i.e.\ by the vacuum \anull.
This illustrates nicely that the vacuum -- due to its enlarged SU(2) symmetry,
represented by the constant gauge transformations -- is `more
singular' than ordinary connections with U(1) holonomy.

{}From the results derived above we learn that $\Lnull/$SU(2) (and the \fmd\
\LG\ in general) is not a very convenient tool for the global description of
reducible connections. Namely, we can show that it is not possible to find
for {\em all\/} reducible connections a representative that is both of the
form \erf{abel} and contained in \Lnull. This is essentially a consequence of
the fact that there is no Gribov effect in abelian gauge theories.
Let $A$ be of the form \erf{abel}, i.e. $A_\mu = \ii\tau a_\mu$.
Then the only gauge copies of $A$ of the form \erf{abel} are
$A_\mu^g = \pm \ii\tau (a_\mu + \dmu \gamma)$. Now if $A$ is contained in
\Lnull, we have $\dmu a^\mu =0$; if $A^g$ is in \Lnull\ as well, it follows
that $\gamma$ has to be harmonic, $\partial^2\gamma = 0$. As all harmonic
functions on $S^4$ are constant, we see that the only elements on the orbit
of $A$ that are contained in $\Lnull$ and are of the form \erf{abel} are
$\pm A$. Therefore the fact that \Lnull\ is bounded shows that, for
any choice of the functions $a_\mu$, the connection $\kappa A$ with
$\kappa$ large enough is not contained in \Lnull\ and that the gauge orbit
through $\kappa A$ does not contain a representative of the form \erf{abel}
in \Lnull.

{}From our results it also follows that the singularities of the configuration
space are conical. More precisely, \calm\ has a `cone over cones' structure.
(A cone over a base space \calb\ is by definition a space which
is diffeomorphic to the direct product $[0, 1] \times \calb$,
where for $t=0$ all points of \calb\ are identified. If the base \calb\ is
endowed with a metric ${\rm d} \Omega$, then the cone has a metric of the form
$ {\rm d}s^2 = {\rm d}t^2 + t^2 {\rm d} \Omega$.
Namely, the set of reducible connections forms the first cone; its tip is the
`vacuum' \anull\ and its base space ${\cal B}_0$ can be described as the real
projective space ${\cal B}_0={\dl RP}^\infty$ that is obtained by the $\zet_2$
identification of antipodal points on the \infdim\ unit sphere
  \be    S := \{ a_\mu | \ixxxx a_\mu a^\mu = 1 \}  \,. \ee

To describe the geometric situation around a connection \bara\ with U(1)
stabilizer we use \bara\ as the background and introduce a neighbourhood $U$ as
we have done in section 4 for general $G$. The intersection $\tilde U$ of $U$
with some linear subspace of $\PSI$ contains locally all other connections with
U(1) stabilizer. Without loss of generality we can assume that $U$ is a direct
sum $U= \tilde U \oplus U^\perp$, where $U^\perp$ is contained in the
orthogonal complement of $\tilde U$. To get the true configuration space we
now have to divide out the residual part of the gauge group, i.e.
$U(1) \cong \cals_{\!\abar} \subset \calg$. This group
acts on $U$ in the following way: it leaves $\tilde U$ pointwise fixed and
freely maps $U^\perp$ on $U^\perp$. To make this more explicit, decompose
any $A\in U$ as $A= \tilde A + A^\perp$ where $\tilde A\in \tilde U$ transforms
as a connection and $A^\perp \in U^\perp$ transforms homogeneously under \calg.
Then a gauge transformation $g \in \calg$ maps $A$ to $A^g = \tilde A +
g^{-1} A^{\perp} g$, where on the last part the action of \calg\ is
homogeneous and free. This shows that around every point of $\tilde U$
again the structure of \calm\ is that of a cone whose tip now lies in
$\tilde U$. The base space of this cone is
  \be {\cal B} = [ S_r^\infty \cap U^\perp ] / \calsa , \ee
where $S_r^\infty$ is a sphere of some radius $r$ in the \infdim\ space
$\Gamma$.
$\cal B$  is a smooth manifold since the action of $\calsa$ on $U^\perp$ is
free. Since any point of $\tilde U$ is itself part of the cone whose tip is
$\anull$, we have indeed identified a `cone over cones' structure of \calm.
\futnote{In a Hamiltonian formulation, for any \YM\ theory a similar
`cones over cones' structure is present \cite{armm,arg.} in the subset
of the phase space that consists of solutions to the constraint equations.
It will be interesting to explore the relation between this cone structure and
the singularity structure of \calm.}

\subsection{Reducible connections on the lattice}

Analogous considerations as for the configuration space of continuum gauge
theories apply when the theory is considered on an arbitrary lattice. In
particular, reducible connections also arise in lattice gauge theories.
 \futnote{Various other aspects of the configuration space of
lattice theories have been treated in the literature. See, for instance,
\cite{maro,vanb3,vava,zwan8}.}
 As it turns out, most of the results of the continuum theory can be
related rather directly to the lattice. For example, the analogue
of the configuration \erf{abel} on an arbitrary lattice reads
  \be  U_{x,y} = \exp(\ii\tau\,f_{x,y})  \,, \labl{lat}
where $U_{x,y}$ denotes the element of SU(2) that is attached to the link
joining the vertices $x$ and $y$,
$\tau$ is an arbitrary element of the Lie algebra su(2),
and where $f$ is an arbitrary real-valued function on the set
of links of the lattice. In the lattice formulation an element of the
gauge group can be described by a map from the vertices $x$ into the
structure group $G$. The group variables attached to the links then transform
according to
  \be U_{x,y} \mapsto g_x^{-1}\, U_{x,y}\, g_y . \ee

One can easily check that any configuration of the form \erf{lat} is invariant
under the U(1) generated by the constant gauge transformations $g_x
= \eE^{\ii\theta\tau} = {\rm const}\,$.
But we can also show the converse: given any configuration $U_{x,y}$
with non-trivial stabilizer of an SU(2) gauge theory on the lattice,
it is gauge equivalent to some configuration of the form \erf{lat}.
Namely, first, the fact that $g$ is in the stabilizer means that
  \be  g^{-1}_x U_{x,y}\, g_y = U_{x,y}  \,. \labl{lafix}
We can extend this formula by iterating it in such a manner that it
applies to any two vertices $x$ and $y$, with $U_{x,y}\equiv U_{x,y}
^{(P_{xy})}$ the transporter for an arbitrary but fixed path $P_{xy}$
joining the vertices. We then fix some reference point \xo,
which, due to the fact that $g$ is non-trivial, we can choose such that
$g_\xO\ne\one$. Then
  \be  g_x = U^{}_{x,\xO} g_\xO^{} U^{-1}_{x,\xO} \labl{gx}
for any lattice point $x$ and any path joining $x$ with \xo; in
particular, while $U_{x,\xO}$ depends on the path chosen, the combination
on the right hand side does not. As $g$ describes a gauge
transformation and hence transforms in the adjoint \rep, \erf{gx} is
just the discrete version of the statement that $g$ is covariantly
constant. Define now the map $\gammA$ on the set of vertices by
$ \gammA_x:= U_{x,\xO}$, and denote by $\tilde U$ the configuration
  \be \tilde U_{x,y} := \gammA_x^{-1} U_{x,y} \gammA_y^{}  \,. \ee
By definition, $\tilde U$ is gauge equivalent to $U$, and
$g_x=\gammA_x g_\xO \gammA_x^{-1}$. In addition, for any $x$ and $y$ we have
  \be \tilde U^{}_{x,y} g_\xO \tilde U^{-1}_{x,y} =
  \gammA_x^{-1} U^{}_{x,y}\gammA_y^{} g_\xO \gammA^{-1}_y U^{-1}_{x,y}
  \gammA^{}_x = \gammA_x^{-1} U^{}_{x,y} g_y U^{-1}_{x,y} \gammA^{}_x =
  \gammA^{-1}_x g_x \gammA^{}_x = g_\xO \,.  \ee
Thus {\em all\/} link variables $\tilde U_{x,y}$ commute with the non-trivial
element $g_\xO$ of SU(2). This shows that they have to be contained in
the U(1) subgroup generated by $g_\xO$, and hence for any configuration
with non-trivial stabilizer there is a gauge equivalent representative of
the form \erf{lat}.

In contrast to the situation in the continuum theory, in a lattice
gauge theory reducible connections do not seem to cause any problems
as long as one works at fixed finite lattice spacing $a$; whether a link
variable is part of a reducible (as in the above example) or an irreducible
configuration seems to be irrelevant.
A crucial difference to the continuum theory is that in the lattice theory
one is not forced to fix a gauge. The partition function is just the
sum over all configurations of link variables. More precisely,
the integration measure is a product of Haar measures of the \findim\
structure group $G$ at each lattice link. With respect to this measure,
reducible connections are a set of measure zero. However, the interest in the
lattice theory comes mostly from the desire to consider it as a regularization
of the continuum theory, and for making contact to observational data
one has ultimately to perform the continuum limit $a\to 0$.
This limit is far from being trivial, and hence it is not at all clear what
the influence of reducible connections might possibly be.
The lattice approach suggests that the measure is concentrated at the
reducible connections (which, nonetheless, are of measure zero) in the
sense that any reducible connection is included in the partition function
with a multiplicity corresponding to its stabilizer. In our opinion this
is an additional hint that these connections should be `blown up' in the
true configuration space.

\vskip12mm
\small

\noindent{\bf Acknowledgement.} \ We are grateful to F.\ Scheck and
P.\ Weisz for interesting discussions. J.\ Fuchs thanks the Institut
f\"ur Theoretische Physik of Heidelberg University for hospitality.

\newpage


   \newcommand{\wb}{\,\linebreak[0]}
   \def\wB                {$\,$\wb}
   \newcommand{\J}[1]     {{{#1}}\vyp}
   \newcommand{\Jj}[1]    {{{#1}}\vyP}
   \newcommand{\JJ}[1]    {{{#1}}\vyp}
   \newcommand{\bi}[1]    {\bibitem{#1}}
   \newcommand{\bI}[2]    {\bibitem{#1}}
   \newcommand{\Bi}[1]    {\bibitem{#1}}
   \newcommand{\BI}[2]    {\bibitem{#1}}
   \newcommand{\Prep}[2]  {preprint {#1}}
   \newcommand{\PRep}[2]  {preprint {#1}}
   \newcommand{\PhD}[2]   {Ph.D.\ thesis (#1)}
   \newcommand{\Erra}[3]  {\,[{\em ibid.}\ {#1} ({#2}) {#3}, {\em Erratum}]}
   \newcommand{\BOOK}[4]  {{\em #1\/} ({#2}, {#3} {#4})}
   \newcommand{\inBO}[7]  {in:\ {\em #1}, {#2}\ ({#3}, {#4} {#5}), p.\ {#6}}
   \newcommand{\inBOnoeds}[6]  {in:\ {\em #1} ({#2}, {#3} {#4}), p.\ {#5}}
   \newcommand{\vyp}[4]   {\ {#1} ({#2}) {#3}}
   \newcommand{\vyP}[3]   {\ ({#1}) {#2}}
   \newcommand{\vypf}[5]  {\ {#1} [FS{#2}] ({#3}) {#4}}
   \def\acam  {Acta\wB Appl.\wb Math.}
   \def\acma  {Acta\wB Math.}
   \def\adma  {Adv.\wb Math.}
   \def\amjm  {Amer.\wb J.\wb Math.}
   \def\amjp  {Amer.\wb J.\wb Phys.}
   \def\amst  {Amer.\wb Math.\wb Soc.\wb Transl.}
   \def\andp  {Ann.\wB der\wB Physik}
   \def\anif  {Ann.\wb Inst.\wB Fou\-rier}
   \def\anip  {Ann.\wb Inst.\wB Poin\-car\'e}
   \def\anma  {Ann.\wb Math.}
   \def\anop  {Ann.\wb Phys.}
   \def\appb  {Acta\wB Phys.\wb Pol. B}
   \def\armi  {Algebraic\wB and\wB Toplogical\wB Theories}
   \def\asen  {Ann.\wb Sci.\wb Ec.\wb Norm.\wb Sup\'er.}
   \def\asnp  {Annali\wB Scuola\wB Norm.\wb Sup.\wB Pisa}
   \def\aspm  {Adv.\wb Stu\-dies\wB in\wB Pure\wB Math.}
   \def\aste  {Ast\'e\-ris\-que}
   \def\bams  {Bull.\wb Amer.\wb Math.\wb Soc.}
   \def\blms  {Bull.\wB London\wB Math.\wb Soc.}
   \def\bsbm  {Bol.\wb Soc.\wb Bras.\wb Math.}
   \def\bsmf  {Bull.\wb Soc.\wb Math.\wB de\wB France}
   \def\busm  {Bull.\wb Sci.\wb Math.}
   \newcommand{\caas}[2] {\inBO{$C^*$-Algebras and their Applications to
              Statistical Mechanics and Quantum Field Theory {\rm
              (Proceedings of the International School of Physics ``Enrico
              Fermi", Course LX)}} {D.\ Kastler, ed.}
              {\NH}{Amsterdam}{1976} {{#1}}{{#2}} }
   \def\cajm  {Ca\-nad.\wb J.\wb Math.}
   \def\camb  {Ca\-nad.\wb Math.\wb Bull.}
   \def\casc  {C.\wb R.\wb Math.\wb Rep.\wb Acad.\wb Sci.\wB Canada}
   \def\clqg  {Class.\wB Quantum\wB Grav.}
   \def\cmsp  {Ca\-nad.\wb Math.\wb Soc.\wb Proc.}
   \def\codi  {Colloque Dixmier}
   \def\coia  {Com\-mun.\wB in\wB Algebra}
   \def\coma  {Con\-temp.\wb Math.}
   \def\comh  {Com\-ment.\wb Math.\wb Helv.}
   \def\comp  {Com\-mun.\wb Math.\wb Phys.}
   \def\core  {Comptes\wB Rendus}
   \def\cnpp  {Com\-ments\wB Nucl.\wb Part.\wb Phys.}
   \def\cpma  {Com\-pos.\wb Math.}
   \def\crap  {C.\wb R.\wb Acad.\wb Sci.\wB Paris}
   \def\ctpa  {Com\-mun.\wb Theor.\wb Phys.\ (Allahabad)}
   \def\ctpb  {Com\-mun.\wb Theor.\wb Phys.\ (Beijing)}
   \def\duke  {Duke\wB Math.\wb J.}
   \def\duki  {Duke\wB Math.\wb J.\ (Int.\wb Mat.\wb Res.\wb Notes)}
   \def\enma  {Enseign.\wb Math.}
   \def\eule  {Europhys.\wb Lett.}
   \def\foph  {Fortschr.\wb Phys.}
   \def\fuaa  {Funct.\wb Anal.\wb Appl.}
   \def\fuek  {Funkcial.\wb Ekvac.}
   \newcommand{\fscp}[2] {\inBO{{\rm Les Houches Summer Session 1988 on}
              Fields, Strings, and Critical Phenomena} {E.\ Br\'ezin and
              J.\ Zinn-Justin, eds.} \NH{Amsterdam}{1989} {{#1}}{{#2}}}
   \newcommand{\ftsm}[2] {\inBO{{\rm Les Houches Summer Session 1982 on}
              Field Theory and Statistical Mechanics} {
              , eds.} \NH{Amsterdam}{1984} {{#1}}{{#2}} \Miss}
   \def\gede  {Geo\-metriae Dedi\-cata}
   \def\haab  {Ab\-handl.\wb Math.\wB Seminar\wB Ham\-burg}
   \def\hepa  {Helv.\wb Phys.\wB Acta}
   \def\ihes  {Publ.\wb Math.\wB I.H.E.S.}
   \def\ijmb  {Int.\wb J.\wb Mod.\wb Phys.\ B}
   \def\ijmc  {Int.\wb J.\wb Mod.\wb Phys.\ C}
   \def\ijmp  {Int.\wb J.\wb Mod.\wb Phys.\ A}
   \newcommand{\ilag}[2] {\inBO{\Infdim\ Lie \A s and Groups {\rm[Adv.\
              Series in Math.\ Phys.\ 7]}} {V.G.\ Kac, ed.} \WS\Si{1989}
              {{#1}}{{#2}}} 
   \def\iljm  {Illinois\wB J.\wb Math.}
   \def\inma  {Invent.\wb math.}
   \def\jajm  {Japan.\wb J.\wb Math.}
   \def\jams  {J.\wb Amer.\wb Math.\wb Soc.}
   \def\jetl  {Sov.\wb Phys.\wB JETP\wB Lett.}
   \def\jetp  {Sov.\wb Phys.\wB JETP}
   \def\jgap  {J.\wb Geom.\wB and\wB Phys.}
   \def\jmsj  {J.\wb Math.\wb Soc.\wB Japan}
   \def\joal  {J.\wB Al\-ge\-bra}
   \def\jodg  {J.\wb Diff.\wb Geom.}
   \def\jofa  {J.\wb Funct.\wb Anal.}
   \def\jomp  {J.\wb Math.\wb Phys.}
   \def\jont  {J.\wB Number\wB Theory}
   \def\jopa  {J.\wb Phys.\ A}
   \def\jopc  {J.\wb Phys.\ C}
   \def\josm  {J.\wb Sov.\wb Math.}
   \def\josp  {J.\wb Stat.\wb Phys.}
   \def\jram  {J.\wB rei\-ne\wB an\-gew.\wb Math.}
   \newcommand{\kniz}[2] {\inBO{The Physics and Mathematics of Strings,
              Memorial Volume for V.G.\ Knizhnik} {L.\ Brink et al., eds.}
              \WS\Si{1990} {{#1}}{{#2}} }
   \def\lemp  {Lett.\wb Math.\wb Phys.}
   \def\lenc  {Lett.\wB Nuovo\wB Cim.}
   \def\leni  {Lenin\-grad\wB Math.\wb J.}
   \def\liam  {Lectures\wB in\wB Applied\wB Math.}
   \def\maan  {Math.\wb Annal.}
   \def\mams  {Memoirs\wB Amer.\wb Math.\wb Soc.}
   \newcommand{\mapx}[2] {\inBO{Mathematical Physics X}
              {K.\ Schm\"udgen, ed.} \SV\Be{1992} {{#1}}{{#2}} }
   \newcommand{\mqft}[2] {\inBO{Modern Quantum Field Theory}
              {S.\ Das, A.\ Dhar, S.\ Mukhi, A.\ Raina, and A.\ Sen, eds.}
              \WS\Si{1991} {{#1}}{{#2}} \Miss }
   \def\maui  {Math.\wB USSR\wB Izv.}
   \def\maze  {Math.\wb Zeitschr.}
   \def\mpla  {Mod.\wb Phys.\wb Lett.\ A}
   \def\mplb  {Mod.\wb Phys.\wb Lett.\ B}
   \def\namj  {Nagoya\wB Math.\wb J.}
   \newcommand{\nicm}[2] {\inBO{IXth International Congress on
              Mathematical Physics} {B.\ Simon, A.\ Truman, and I.M.\ Davis,
              eds.} {Adam Hilger}{Bristol}{1989} {{#1}}{{#2}} } 
   \def\npbf  {Nucl.\wb Phys.\ B\vypf}
   \def\npbp  {Nucl.\wb Phys.\ B (Proc.\wb Suppl.)}
   \newcommand{\nqft}[2] {\inBO{ 
              Nonperturbative Quantum Field Theory} {G.\ 't Hooft, A.\
              Jaffe, G.\ Mack, P.K.\ Mitter, and R.\ Stora, eds.}
              \PL\NY{1988} {{#1}}{{#2}} }
   \newcommand{\nspq}[2] {\inBO{ 
              New Symmetry Principles in Quantum Field Theory}
              {J.\ Fr\"ohlich et al., eds.} \PL\NY{1992} {{#1}}{{#2}}}
   \def\nuci  {Nuovo\wB Cim.}
   \def\nupb  {Nucl.\wb Phys.\ B}
   \def\osjm  {Osaka\wB J.\wb Math.}
   \def\paaa  {Proc.\wb Amer.\wb Acad.\wB Arts\wB Sci.}
   \def\pams  {Proc.\wb Amer.\wb Math.\wb Soc.}
   \def\pcps  {Proc.\wB Cam\-bridge\wB Philos.\wb Soc.}
   \def\phla  {Phys.\wb Lett.\ A}
   \def\phlb  {Phys.\wb Lett.\ B}
   \def\phle  {Phys.\wb Lett.}
   \def\phrb  {Phys.\wb Rev.\ B}
   \def\phrd  {Phys.\wb Rev.\ D}
   \def\phre  {Phys.\wb Rev.}
   \def\phrl  {Phys.\wb Rev.\wb Lett.}
   \def\phsc  {Physica\wB Scripta}
   \def\phya  {Physica\ A}
   \newcommand{\phgt}[2] {\inBO{Physics, Geometry, and Topology}
              {H.C.\ Lee, ed.} \PL\NY{1990} {{#1}}{{#2}} }
   \def\phyd  {Physica\ D}
   \def\pajm  {Pa\-cific\wB J.\wb Math.}
   \def\pkna  {Proc.\wb Kon.\wb Ned.\wb Akad.\wb Wetensch.}
   \def\plms  {Proc.\wB Lon\-don\wB Math.\wb Soc.}
   \def\pnas  {Proc.\wb Natl.\wb Acad.\wb Sci.\wb USA}
   \def\prep  {Phys.\wb Rep.}
   \def\prja  {Proc.\wB Japan\wB Acad.}
   \def\prsa  {Proc.\wb Roy.\wb Soc.\wB Ser.$\,$A}
   \def\prtp  {Progr.\wb Theor.\wb Phys.}
   \def\pspm  {Proc.\wb Symp.\wB Pure\wB Math.}
   \def\ptps  {Progr.\wb Theor.\wb Phys.\wb Suppl.}
   \def\ptrs  {Phil.\wb Trans.\wb Roy.\wb Soc.\wB Lon\-don}
   \def\remp  {Rev.\wb Mod.\wb Phys.}
   \def\rpmp  {Rep.\wb Math.\wb Phys.}
   \def\rimk  {RIMS\wB Kokyuroku}
   \def\rims  {Publ.\wB RIMS}
   \def\rmap  {Rev.\wb Math.\wb Phys.}
   \def\rums  {Russ.\wb Math.\wb Surveys}
   \newcommand{\sagt}[2] {\inBO{{\rm Proceedings of the} Symposium on
              Anomalies, Geometry, Topology} {W.A.\ Bardeen and A.R.\
              White, eds.} \WS\Si{1985} {{#1}}{{#2}} }
   \def\sebo  {S\'emi\-naire\wB Bour\-baki}
   \def\siaa  {SIAM\wB J.\wb Alg.\wb Disc.\wb Meth.}
   \def\siam  {SIAM\wB J.\wb Math.\wb Anal.}
   \newcommand{\sisv}[2] {\inBO{{\rm Proceedings of the International
              Symposium on} Symmetries in Science V} {B.\ Gru\-ber, ed.}
              \PL\NY{1991} {{#1}}{{#2}} }
   \def\sjnp  {Sov.\wb J.\wb Nucl.\wb Phys.}
   \def\slnm  {Sprin\-ger Lecture Notes in Mathematics}
   \def\slnp  {Sprin\-ger Lecture Notes in Physics}
   \def\somd  {Sov.\wb Math.\wb Dokl.}
   \def\sopd  {Sov.\wb Phys.\wb Dokl.}
   \def\sopu  {Sov.\wb Phys.\wb Usp.}
   \def\spaw  {Sitzungs\-ber.\wb Preuss.\wb Akad.\wb Wiss.}
   \def\ssra  {Sov.\wb Sci.\wb Rev.\ A}
   \def\ssrc  {Sov.\wb Sci.\wb Rev.\ C}
   \newcommand{\stri}[2] {\inBO{Strings 89{\rm, Proceedings of the
              College Station String Workshop 1989}} {R.\ Arnowitt et
              al., eds.} \WS\Si{1990} {{#1}}{{#2}} }
   \newcommand{\Suse} [2] {\inBO{The Algebraic Theory of Superselection
              Sectors.\ Introduction and Recent Results} {D. Kastler,
              ed.} \WS\Si{1990} {{#1}}{{#2}} }
   \def\tams  {Trans.\wb Amer.\wb Math.\wb Soc.}
   \def\thmp  {Theor.\wb Math.\wb Phys.}
   \def\tojm  {Tokyo\wB J.\wb Math.}
   \def\tomj  {Tohoku\wB Math.\wb J.}
   \def\topo  {Topology}
   \newcommand{\voms}[2] {\inBO{Vertex Operators in Mathematics and
              Physics {\rm [M.S.R.I.\ publication No.\ 3]}} {J.\ Lepowsky,
              S.\ Mandelstam, and I.M.\ Singer, eds.} \SV\NY{1985}
              {{#1}}{{#2}} } 
   \def\zfpb  {Z.\wb Phy\-sik B}
   \def\zfpc  {Z.\wb Phy\-sik C}
   \def\zfph  {Z.\wb Phy\-sik}
   \def\AMS    {{American Mathematical Society}}
   \def\AP     {{Academic Press}}
   \def\AW     {{Addi\-son\hy Wes\-ley}}
   \def\BC     {{Ben\-jamin\,/\,Cum\-mings}}
   \def\BIR    {{Birk\-h\"au\-ser}}
   \def\CUP    {{Cambridge University Press}}
   \def\CUPC   {{Cambridge University Press}}
   \def\DP     {{Dover Publications}}
   \def\GB     {{Gordon and Breach}}
   \def\JW     {{John Wiley}}
   \def\KLU    {{Kluwer Academic Publishers}}
   \def\MD     {{Marcel Dekker}}
   \def\MGH    {{McGraw\,\hy\,Hill}}
   \def\NH     {{North Holland Publishing Company}}
   \def\OUP    {{Oxford University Press}}
   \def\PL     {{Plenum}}
   \def\PUP    {{Princeton University Press}}
   \def\SV     {{Sprin\-ger Verlag}}
   \def\WI     {{Wiley Interscience}}
   \def\WS     {{World Scientific}}
   \def\Be     {{Berlin}}
   \def\Ca     {{Cambridge}}
   \def\NY     {{New York}}
   \def\pR     {{Princeton}}
   \def\PR     {{Providence}}
   \def\Si     {{Singapore}}

\end{document}